\documentclass[onecolumn,showpacs,preprintnumbers,amsmath,amssymb]{revtex4}

\usepackage{graphicx}
\usepackage{graphics}
\usepackage{dcolumn}
\usepackage{bm}

\def\BEq{\begin{equation}}
\def\EEq{\end{equation}}
\def\BEqA{\begin{eqnarray}}
\def\EEqA{\end{eqnarray}}
\def\BEn{\begin{enumerate}}
\def\EEn{\end{enumerate}}
\def\BWT{\begin{widetext}}
\def\EWT{\end{widetext}}
\def\a{\alpha}

\def\d{\delta}

\def\ket{\rangle}

\begin{document}


\title{Maximally entangling tripartite protocols for Josephson phase qubits}
\author{Andrei Galiautdinov}
 \email{ag@physast.uga.edu}
\affiliation{
Department of Physics and Astronomy,
University of Georgia, Athens, Georgia
30602
}
\author{John M. Martinis}
 \email{martinis@physics.ucsb.edu}
\affiliation{
Department of Physics, University of California, Santa Barbara, 
California 93106}

\date{\today}

\begin{abstract}

We introduce a suit of simple entangling protocols for 
generating  tripartite GHZ and W states in systems with anisotropic 
exchange interaction $g\left(XX+YY\right)+\tilde{g}ZZ$.
An interesting example is provided by macroscopic entanglement 
in Josephson phase qubits with capacitive ($\tilde{g}=0$) and 
inductive ($0<|\tilde{g}/g| < 0.1$) couplings.

\end{abstract}

\pacs{03.67.Bg, 03.67.Lx, 85.25.-j}    

\maketitle



\section{Introduction}

Superconducting circuits with Josephson junctions have attracted 
considerable attention as promising candidates for scalable 
solid-state quantum computing architectures. The story began in 
the early 1980's, when Tony Leggett made a remarkable prediction 
that under certain experimental conditions the macroscopic variables 
describing such circuits could exhibit a characteristically quantum 
behavior
\cite{LEGGETT80}. Several years later such 
behavior was unambiguously observed in a series of tunneling 
experiments by Devoret et al. \cite{DEVORETETAL85}, Martinis et al.
\cite{MARTINISETAL87}, and Clarke et al.
\cite{CLARKEETAL88}. It was eventually realized 
that due to their intrinsic anharmonicity, the ease of manipulation, 
and relatively long coherence times
\cite{VION_LongCoherTimes02}, the metastable 
macroscopic quantum states of the junctions could be used as the 
states of the qubits. That idea had recently been supported by 
successful experimental demonstrations of Rabi oscillations 
\cite{NAKAMURA_RabiOsc97}, high-fidelity
state preparation and measurement \cite{WALLRAFF_SingleShotMeas05, 
SIMMONDS_SingleShotMeas04, 
COOPER_SingleShotMeas04, CLAUDON_SingleShotMeas04, 
KATZ_SingleShotMeas06, MCDERMOTT_SingleShotMeas05}, and various 
logic gate operations 
\cite{SIMMONDS_SingleShotMeas04, COOPER_SingleShotMeas04, 
CLAUDON_SingleShotMeas04, KATZ_SingleShotMeas06, JOHNSONETAL2003}. 
Further progress in developing a workable quantum
computer will depend on the architecture's ability to generate 
various multiqubit entangled states that form the basis for many 
important information processing algorithms \cite{NIELSEN}.

In this paper we develop several {\it single-step} entangling protocols 
suitable for generating maximally entangled quantum states in 
tripartite systems with pair-wise coupling 
$g\left(XX+YY\right)+\tilde{g}ZZ$. We base our 
approach on the idea that implementing symmetric states may conveniently 
be done by symmetrical control of all the qubits in the system. This bears 
a resemblance to approaches routinely used in digital electronics: while 
an arbitrary
gate (for example, a 3-bit gate) can be made from a collection of NAND gates, 
it is often convenient to use more complicated designs with three input 
logic gates to make the needed gate faster and/or smaller. 

The protocols developed in this paper may be directly applied to virtually any of
the currently known superconducting qubit architectures, two of which will be 
mentioned here. The first architecture is based on capacitively coupled
current-biased (CBJJ) Josephson junctions \cite{JOHNSONETAL2003, BLAIS2003,
MCDERMOTT_SingleShotMeas05}
whose dynamics is governed by the circuit Hamiltonian 
\BEq
H_1 = \left(p_1^2+p_2^2+\kappa p_1p_2\right)/2m
+ (\hbar/2e)\sum_{i=1}^{2} \left[-I_0 \cos \phi_i - I_i\phi_i\right],
\EEq
with
$p_i = m\dot{\phi}_i$, $ m = (\hbar/2e)^2\left(C+C_{\rm int}\right)$,
$\kappa = 2C_{\rm int}/(C+C_{\rm int})$.
The other architecture involves inductively coupled flux-biased 
(FBJJ) junctions
\cite{YOU_NAKAMURA_NORI_2005}.
It 
is described by the Hamiltonian
(for small $\Upsilon$, see Ref. \cite{MYCNOTPAPER1} for details)
\BEq
H_2 = \left(p_1^2+p_2^2\right)/2m
+ (\hbar/2e)\sum_{i=1}^{2} \left[-I_0 \cos \phi_i + (eE_0/\hbar)
\left(\phi_i - 2\pi\Phi_i/\Phi_{\rm sc}\right)^2\right]
+ \Upsilon E_0
\left(\phi_1 
-2\pi\Phi_i/\Phi_{\rm sc}\right)
\left(\phi_2  
-2\pi\Phi_i/\Phi_{\rm sc}\right),
\EEq
with
$p_i = m\dot{\phi}_i$, $m = (\hbar/2e)^2C$, $\omega_0 = 1/\sqrt{LC}$,
$E_0= \hbar^2\omega_0^2/2E_C$, $E_C = (2e)^2/2C$, 
$\Phi_{\rm sc}=h/2e$, $\Upsilon = M/L$.
When reduced to computational subspace, in the rotating wave 
approximation, these Hamiltonians become
\BEq
H_{\rm RWA} = (1/2)\left[
\vec{\Omega}_1 \cdot \vec{\sigma}_1 + \vec{\Omega}_2\cdot\vec{\sigma}_2
+ g\left(\sigma_x^1 \sigma_x^2+\sigma_y^1 \sigma_y^2\right)
				+ \tilde{g}\sigma_z^1\sigma_z^2\right],
\EEq
with $\tilde{g}=0$ (momentum-momentum coupling)
and $0< \tilde{g}/g<0.1$ \cite{GELLER_private_comm}
(position-position coupling), respectively. For typical superconducting 
qubits, the RWA requirements are well 
satisfied: the level splittings are usually around 
$\omega/2\pi \simeq 10$ GHz and the Rabi frequencies 
$\Omega$ needed to implement various logic gates 
\cite{MYCNOTPAPER1, MYCNOTPAPER3} are on the order of the coupling 
constant 
$g/2\pi \lesssim 100$ MHz. Thus, $\Omega/\omega \sim 10^{-2}$, 
as required.

\section{The GHZ protocol}
\label{sec:GHZprotocol}

\subsection{Triangular coupling scheme}

In the rotating frame in the absence of coupling, the computational 
basis states
$|000\ket$, $|001\ket$, $|010\ket$, $|100\ket$, $|011\ket$, $|101\ket$, 
$|110\ket$, $|111\ket$ have the same effective energy $E_{\rm eff}=0$. 
The pair-wise coupling,
\BEq
\label{eq:Hint}
H_{\rm int} =
 (1/2){\sum^3_{i, j = 1}}
 g
 \left(
 \sigma_x^i\sigma_x^j+\sigma_y^i \sigma_y^j\right)
 +\tilde{g}\sigma_z^i\sigma_z^j
=
\begin{pmatrix} 
3\tilde{g}/2&    &    &    &    &    &    &   
  \cr
& -\tilde{g}/2&  g&    g&   &    &  &
   \cr
&  g&  -\tilde{g}/2&   g&   &    &     &   
   \cr
&  g&    g&    -\tilde{g}/2&    &   &   &  
   \cr
 &&&&   -\tilde{g}/2&    g&    g     &
   \cr
 &&&&    g&    -\tilde{g}/2&   g    & 
   \cr
 &&&&    g&   g&    -\tilde{g}/2   &    
   \cr   
   & &&&&&&&3\tilde{g}/2 
\cr 
\end{pmatrix},
\EEq
(empty matrix elements are zero), {\it partially lifts} the degeneracy, 
which 
results in the energy spectrum
\BEq
\label{eq:HintDIAG}
E_{\rm int} =
\left\{
3\tilde{g}/2, \, 3\tilde{g}/2, \, 
2g-\tilde{g}/2, \, 2g-\tilde{g}/2, \,
-\left(g+\tilde{g}/2\right), \, -\left(g+\tilde{g}/2\right), \, 
-\left(g+\tilde{g}/2\right), \, -\left(g+\tilde{g}/2\right)  
\right\},
\EEq
with the corresponding ${\cal H}$-eigenbasis
\BEq
{\cal H}_{\rm GHZ}\bigoplus {\cal H}_{\rm W}\bigoplus {\cal H}_{\rm rest}
\equiv  
\left\{|000\ket \oplus |111\ket\right\}\bigoplus 
\left\{|{\rm W}\ket \oplus |{\rm W}'\ket\right\}\bigoplus
\left\{|\Psi_1\ket \oplus |{\Psi_1}'\ket \oplus |\Psi_2\ket \oplus  
|{\Psi_2}'\ket\right\} ,
\EEq
where
\BEqA
&& |{\rm W}\ket = (|100\ket+|010\ket+|001\ket)/\sqrt{3},\quad
|{\rm W}'\ket = (|011\ket+|101\ket+|110\ket)/\sqrt{3},\nonumber \\
&& |\Psi_1\ket =(|100\ket-|010\ket)/\sqrt{2},\quad
|{\Psi_1}'\ket = (|011\ket-|101\ket)/\sqrt{2},\nonumber \\
&&|\Psi_2\ket = (|100\ket + |010\ket - 2\,|001\ket)/\sqrt{6} , \quad
|{\Psi_2}'\ket = (|011\ket + |101\ket - 2\,|110\ket)/\sqrt{6}.
\EEqA

Since the coupling does not cause transitions within each of the 
degenerate subspaces (nor does it cause transitions between different 
such subspaces), it is impossible to generate the $|{\rm GHZ}\ket 
= \left(|000\ket + |111\ket\right)/\sqrt{2}$ 
state \cite{GHZ89}
from the ground state $|000\ket$ by direct application of $H_{\rm int}$. 
Instead, we must first 
bring the $|000\ket$ state out of the ${\cal H}_{\rm GHZ}$ subspace by, 
for example, subjecting it to 
a local rotation $R_1$ in such a way as to produce a state 
 $|\psi\ket$ that has both $|000\ket$ and $|111\ket$ components. That 
 is only possible if {\it all} one-qubit amplitudes 
 $\alpha_1, \dots, \beta_3$ in the resulting product state
$|\psi\ket = R_1|000\ket = \left(\alpha_1|0\ket +\beta_1 
|1\ket \right)
\left( \alpha_2|0\ket +\beta_2 |1\ket\right)\left( \alpha_3|0\ket 
+\beta_3 |1\ket\right)
$
are chosen to be nonzero, which means that in the computational basis 
the state $|\psi\ket$ will have eight nonzero components. 

We now notice that in the ${\cal H}$-basis, the three-qubit rotations 
\BEqA
X_{\theta} &=& X^{(3)}_{\theta}X^{(2)}_{\theta}X^{(1)}_{\theta} =
\begin{pmatrix}
c^3 &  is^3 & -i\sqrt{3}sc^2 & -\sqrt{3}cs^2
\cr
is^3 & c^3 & -\sqrt{3}cs^2 & -i\sqrt{3}sc^2 
\cr
-i\sqrt{3}sc^2 & -\sqrt{3}cs^2 &  c(1-3s^2) &        is(1-3c^2) 
\cr
-\sqrt{3}cs^2 & -i\sqrt{3}sc^2 & is(1-3c^2) &  c(1-3s^2) 
\cr
\end{pmatrix}
\oplus
\begin{pmatrix}
c &   is 
\cr
is &  c
\cr
\end{pmatrix}
\oplus
\begin{pmatrix}
 c &    is
\cr
is & c
\cr
\end{pmatrix},
\nonumber \\
Y_{\theta} &=& Y^{(3)}_{\theta}Y^{(2)}_{\theta}Y^{(1)}_{\theta} =
\begin{pmatrix}
c^3 & -s^3 & -\sqrt{3}sc^2 &  \sqrt{3}cs^2 
 \cr
s^3 & c^3 &  \sqrt{3}cs^2 &  \sqrt{3}sc^2 
\cr
\sqrt{3}sc^2 &  \sqrt{3}cs^2 &   c(1-3s^2) &    s(1-3c^2) 
\cr
\sqrt{3}cs^2 & -\sqrt{3}sc^2 &  -s(1-3c^2) &  c(1-3s^2)
\cr
\end{pmatrix}
\oplus
\begin{pmatrix}
 c & s 
\cr
-s & c 
\cr
\end{pmatrix}
\oplus
\begin{pmatrix}
  c & s
\cr
 -s & c
\cr
\end{pmatrix},
\EEqA
where
$Y^{(k)}_{\theta} = \exp\left(-i\theta \sigma_y^k/2\right)$,
$X^{(k)}_{\theta} = \exp\left(-i\theta \sigma_x^k/2\right)$, $k = 1,2,3$,
are block-diagonal,
with $c \equiv \cos(\theta/2)$ and $s \equiv \sin(\theta/2)$.
For $\theta = \pi/2$, the corresponding $4\times 4$ blocks acting on the
${\cal H}_{\rm GHZ} \bigoplus {\cal H}_{\rm W}$ subspace are
\BEq
X^{(4\times 4)}_{\pi/2} =
\frac{1}{\sqrt{8}}
\begin{pmatrix}
1 &  i & -i\sqrt{3} & -\sqrt{3}
\cr
i& 1 & -\sqrt{3}& -i\sqrt{3} 
\cr
-i\sqrt{3} & -\sqrt{3} &  -1 & -i
\cr
-\sqrt{3} & -i\sqrt{3} & -i &  -1
\cr
\end{pmatrix}, 
\quad
Y^{(4\times 4)}_{\pi/2}=
\frac{1}{\sqrt{8}}
\begin{pmatrix}
1 & -1 & -\sqrt{3} &  \sqrt{3}
 \cr
1 & 1 &  \sqrt{3} &  \sqrt{3}
\cr
\sqrt{3} &  \sqrt{3} &   -1 &  -1 
\cr
\sqrt{3} & -\sqrt{3} &  1 & -1
\cr
\end{pmatrix}.
\EEq
This shows that $Y_{\pi/2}$ provides a convenient choice for $R_1$. 
We may thus start by generating the so-called symmetric state,
\BEq
\label{eq:symmetricSTATE}
{|\psi\ket}_{\rm sym} 
= Y_{\pi/2} |000\ket
=
(1/2)\left(
|{\rm GHZ}\ket
+ \sqrt{3/2}\; \left(|{\rm W}\ket +|{\rm W}'\ket\right)
\right)  
\in {\cal H}_{\rm GHZ} \bigoplus {\cal H}_{\rm W}.
\EEq
The entanglement is then performed by acting on ${|\psi\ket}_{\rm sym}$
with $U_{\rm int}=\exp\left(-iH_{\rm int}t\right)$, thus inducing a 
phase difference between the GHZ and W+${\rm W}'$ components
(this step works only for $g\neq \tilde{g}$, see Section 
\ref{sec:propertiesISOTROPIC}),
\BEq
\label{eq: entanglement GHZ}
U_{\rm int} Y_{\pi/2} |000\ket
=\left(e^{-i\a}/2\right)\left(
|{\rm GHZ}\ket
+ e^{-i\d} \sqrt{3/2}\; \left(|{\rm W}\ket +|{\rm W}'\ket\right)
\right),
\quad \a = \left(3\tilde{g}/2\right)t, 
\quad \d = 2 \left(g-\tilde{g}\right)t.
\EEq 

To transform to the desired GHZ state, we first diagonalize the 
$X^{(4\times 4)}_{\pi/2}$ and $Y^{(4\times 4)}_{\pi/2}$ operators 
to get the unimodular spectra 
\BEq
\label{eq:XYspectra}
\lambda_X = \left\{-e^{i(\pi/4)}, -e^{-i(\pi/4)}, e^{-i(\pi/4)}, 
e^{i(\pi/4)}  \right\}, \quad
\lambda_Y = \left\{-e^{-i(\pi/4)}, -e^{i(\pi/4)}, e^{i(\pi/4)}, 
e^{-i(\pi/4)}  \right\},
\EEq
corresponding to the ${\cal X}$- and ${\cal Y}$-eigenbases,
${\cal X}=
\begin{pmatrix}
|X_1\ket & |X_2\ket & |X_3\ket & |X_4\ket
\end{pmatrix} \equiv Y^{(4\times 4)}_{\pi/2}$, 
${\cal Y} =
\begin{pmatrix}
|Y_1\ket & |Y_2\ket & |Y_3\ket & |Y_4\ket
\end{pmatrix} \equiv X^{(4\times 4)}_{\pi/2}$,
which are formed by the columns of $Y^{(4\times 4)}_{\pi/2}$ and 
$X^{(4\times 4)}_{\pi/2}$.
Using the ${\cal X}$-basis, we notice that both states
\BEq
|{\rm GHZ}\ket = \frac{|X_1\ket+\sqrt{3}|X_4\ket}{2},
\quad
U_{\rm int} Y_{\pi/2} |000\ket
=\frac{e^{-i\a}}{2}\left(
\frac{1+3e^{-i\d}}{2}|X_1\ket+\frac{1-e^{-i\d}}{2}\sqrt{3}|X_4\ket 
\right),
\EEq 
belong to the same two-dimensional (nondegenerate) ${\cal X}$-subspace 
spanned by $|X_1\ket \oplus |X_4\ket$.
Therefore, by performing an additional $X_{\pi/2}$ rotation we can transform
$U_{\rm int} Y_{\pi/2} |000\ket$ to
\BEq
X_{\pi/2} U_{\rm int} Y_{\pi/2}|000\ket = e^{-i\a}e^{i(\pi/4)}|{\rm GHZ}\ket,
\EEq
provided the entangling time is set to give $|\d|=\pi$, or,
$t_{\rm GHZ} = \pi/2|g-\tilde{g}|$.
Any other GHZ state $(|000\ket + e^{i\phi}|111\ket)/\sqrt{2}$ can 
be made out of the ``standard'' GHZ state by a $Z$-rotation applied 
to {\it one} of the qubits, as usual.

The protocol may be compared to controlled-NOT logic gate implementations
\cite{MYCNOTPAPER1, MYCNOTPAPER3} that used various sequences
${\rm CNOT} = e^{-i(\pi/4)}R_2 U_{\rm CNOT}R_1$,
with (entangling) times
$t_{\rm CNOT} = T\pi/2g$, $1\leq T < 1.6$. Thus, for $\tilde{g}=0$, 
the entangling operation proposed here will be 
of same duration as the fastest possible CNOT.

We conclude this section by noting that in its present form the GHZ 
protocol {\it cannot} be used to generate the W state. This can be 
seen by writing
$|{\rm W}\ket = 
\left(
\sqrt{3}\left(|X_1\ket + |X_2\ket\right)
-(|X_3\ket + |X_4\ket)
\right)/\sqrt{8}$,
which shows that our $XU_{\rm int}Y$ sequence does not result in a W
since the final $X_{\pi/2}$ rotation cannot eliminate the $|X_2\ket$ 
and $|X_3\ket$ components. 
Also,
\BEq
\label{eq:W_Y}
|{\rm W}\ket =
\left(
\sqrt{3}\left(i |Y_1\ket -|Y_2\ket\right)
-(|Y_3\ket - i|Y_4\ket)
\right)/\sqrt{8}, 
\EEq
and
\BEq
\label{eq:W_Y_no}
Y_{\pi/2} U_{\rm int} Y_{\pi/2} |000\ket
=e^{-i\a}\left(
\frac{1-3e^{-i\d}}{2}\left( i|Y_1\ket -|Y_2\ket \right)
-\frac{\sqrt{3}\left(1+e^{-i\d}\right)}{2}\left(|Y_3\ket-i|Y_4\ket\right) 
\right)/\sqrt{8},
\EEq
and thus no choice of $\d$ will work for the $Y U_{\rm int} Y$ sequence
either.

\subsection{Linear coupling scheme}

In the case of linear coupling, say, $1\leftrightarrow 2$ and
$2\leftrightarrow 3$, the interaction Hamiltonian is given by
\BEq
\label{eq:Hint_2couplers}
H_{\rm int}
=
\begin{pmatrix} 
\tilde{g} &   &    &    &    &   &    &   \cr
   &  0  &  g  & 0 &   &   &   &   \cr
    &   g & -\tilde{g}&   g &    &    &   &   \cr
   &   0&   g&  0 &   &   &   &   \cr
   &   &   &   & 0  &   g& 0 &   \cr
   &   &   &   &   g& -\tilde{g}&   g&   \cr
   &   &   &   &  0 &   g&  0 &   \cr
   &   &   &   &   &   &   &  \tilde{g}\cr 
\end{pmatrix},
\quad
E_{\rm int} =
\left\{
\tilde{g}, \, \tilde{g}, \, 
\epsilon^{(+)}, \, 
\epsilon^{(+)}, \,
\epsilon^{(-)}, \, 
\epsilon^{(-)}, \,
0, \, 0
\right\},
\quad 
\varepsilon^{(\pm)} = \pm \sqrt{2g^2+\left(\tilde{g}/2\right)^2}-
\tilde{g}/2,
\EEq
with eigenbasis
\BEqA
&& |000\ket, \quad |111\ket, \nonumber \\
&& {|{\rm W}\ket}^{(+)} = C^{(+)}
\left(|001\ket+ (\epsilon^{(+)}/g)|010\ket+|001\ket\right),
\quad
{|{\rm W}'\ket}^{(+)} = 
C^{(+)}
\left(|011\ket+ (\epsilon^{(+)}/g)|101\ket+|110\ket\right),
\nonumber \\
&& 
{|{\rm W}\ket}^{(-)} = C^{(-)}
\left(|001\ket+ (\epsilon^{(-)}/g)|010\ket+|001\ket\right),
\quad
{|{\rm W}'\ket}^{(-)} = 
C^{(-)}
\left(|011\ket+(\epsilon^{(-)}/g)|101\ket+|110\ket\right),
\nonumber \\
&&
|\Psi\ket = (|001\ket - |100\ket)/\sqrt{2}  ,\quad 
|{\Psi}'\ket  = (|011\ket - |110\ket)/\sqrt{2},
\EEqA
where $C^{(\pm)}$ are normalizing constants.
We have, 
\BEq
|{\rm W}\ket = A^{(+)}{|{\rm W}\ket}^{(+)}
+ A^{(-)}{|{\rm W}\ket}^{(-)},
\quad
A^{(+)}=  
\frac{-\epsilon^{(-)}+g}{\epsilon^{(+)}-\epsilon^{(-)}} 
\left(\frac{1}{C^{(+)}}\right),
\quad
A^{(-)}= 
\frac{\epsilon^{(+)}-g}{\epsilon^{(+)}-\epsilon^{(-)}}
\left(\frac{1}{C^{(-)}}\right),
\EEq 
and similarly for $|{\rm W}'\ket$. Our GHZ sequence then leads to 
the entangled state
\BEq
U_{\rm int} Y_{\pi/2} |000\ket
=\left(e^{-i\a}/2\right)
\left(
|{\rm GHZ}\ket 
+\sqrt{3/2} \left(
e^{-i\d^{(+)}}A^{(+)} \left[{|{\rm W}\ket}^{(+)}
+{|{\rm W'}\ket}^{(+)}\right]
+ e^{-i\d^{(-)}}A^{(-)}
\left[{|{\rm W}\ket}^{(-)}+{|{\rm W'}\ket}^{(-)}\right]\right)
\right),
\EEq
with
$\a = \tilde{g}t$, $\d^{(\pm)} = (\epsilon^{(\pm)}-\tilde{g})t$.
Since $t>0$, in order for the $X_{\pi/2}$ post-rotation to give a GHZ,
we must restrict coupling to $\tilde{g}=0$
and set the entangling time to $t_{\rm GHZ}=\pi/\sqrt{2}|g|$.
An alternative GHZ implementation for superconducting qubit systems with 
capacitive coupling has recently been considered in Refs. 
\cite{WEI06, MATSUO07}. There, individual qubits were conditionally 
operated upon one at a time.

\section{The W protocol}
\label{sec:Wprotocol}

We now turn to the W protocol. Eq. (\ref{eq:W_Y_no}) suggests that 
control sequence $Y U_{\rm int} Y$ may still give a W, provided a 
proper adjustment of $i|Y_1\ket-|Y_2\ket$ and $|Y_3\ket-i|Y_4\ket$ 
amplitudes is made by a physically acceptable change of system's 
Hamiltonian. In the context of Josephson phase qubits such modification 
can be achieved by adding local Rabi term(s) to $H_{\rm int}$,
for instance,
\BEq
H^{\Omega}_{\rm int} = \left(\Omega/2\right)\left(\sigma_x^1+\sigma_x^2
+\sigma_x^3\right)
+H_{\rm int} =
\frac{1}{2}
\begin{pmatrix} 
 3\tilde{g}&  \Omega&  \Omega& \Omega &   &   &    &   \cr
 \Omega& -\tilde{g}&   2g& 2g & \Omega&   \Omega&  &  \cr
    \Omega&   2g& -\tilde{g}& 2g& \Omega&  &  \Omega&    \cr
  \Omega & 2q&  2g&  -\tilde{g}&  &   \Omega&   \Omega&   \cr
  & \Omega& \Omega&  &  -\tilde{g}&  2g&  2g& \Omega \cr
   &  \Omega&  &  \Omega& 2g& -\tilde{g}& 2g&  \Omega\cr
   & &  \Omega& \Omega& 2g& 2g&  -\tilde{g}&  \Omega\cr
   &  &  & & \Omega &   \Omega&  \Omega& 3\tilde{g}\cr 
\end{pmatrix}.
\EEq
The energy spectrum then becomes
$ E^{\Omega}_{\rm int} = 
\left\{
\epsilon^{(+)} \pm \chi^{(+)}, \, 
\epsilon^{(-)} \pm \chi^{(-)},\,
-\epsilon^{(+)}, \,
-\epsilon^{(+)}, \,
-\epsilon^{(-)}, \,
-\epsilon^{(-)}
\right\}$,
with
$\epsilon^{(\pm)} = g+ \tilde{g}/2 \pm \Omega/2$,
$\chi^{(\pm)}= \sqrt{(g-\tilde{g})^2 \pm (g-\tilde{g})\Omega + \Omega^2 }$.
The (first two) eigenvectors are
\BEq
|\Phi^{(+)}_{1,2}\ket = C_{1,2}^{(+)}
\left( 
\left[-1
-(2/\Omega)\left(
g-\tilde{g}
\mp \chi^{(+)}
\right)
\right]
|{\rm GHZ}\ket
+ \sqrt{3/2}\; 
\left(|{\rm W}\ket +|{\rm W}'\ket\right)
\right),
\EEq
with normalizing constants $C^{(+)}_k$, $k=1,2$. After some algebra we find:
\BEq
\label{eq: entanglement W}
U^{\Omega}_{\rm int} Y_{\pi/2} |000\ket
=
 e^{-i\a}/(4\sqrt{2}\chi^{(+)})
\left[(A/\Omega)\left( i e^{i(\pi/4)}|Y_1\ket - e^{-i(\pi/4)}|Y_2\ket \right)
+(\sqrt{3}B/\Omega)\left(e^{-i(\pi/4)}|Y_3\ket - ie^{i(\pi/4)}|Y_4\ket\right) 
\right],
\EEq
where
\BEqA
A &=& 
\left(g-\tilde{g}+\Omega+\chi^{(+)}\right)
\left(g-\tilde{g}+2\Omega-\chi^{(+)}\right)
-e^{-i\d}\left(g-\tilde{g}+\Omega-\chi^{(+)}\right)
\left(g-\tilde{g}+2\Omega+\chi^{(+)}\right),
\nonumber \\
B &=&
\left(g-\tilde{g}+\Omega+\chi^{(+)}\right)
\left(g-\tilde{g}-\chi^{(+)}\right)
-e^{-i\d}\left(g-\tilde{g}+\Omega-\chi^{(+)}\right)
\left(g-\tilde{g}+\chi^{(+)}\right),
\EEqA 
and $\a = \left(\epsilon^{(+)} + \chi^{(+)}\right)t$, 
$\d = -2\chi^{(+)}t$.
It is straightforward to verify that additional $Y_{\pi/2}$ 
rotation applied to this state produces a W (see Eqs. 
(\ref{eq:XYspectra}), (\ref{eq:W_Y})), 
\BEq
Y_{\pi/2} U^{\Omega}_{\rm int} Y_{\pi/2}|000\ket 
= \left[-{\rm sgn}\left(g-\tilde{g}\right)\right]
e^{-i\a}|{\rm W}\ket,
\EEq
provided we set
$t_{\rm W}=\pi/\sqrt{3}|g-\tilde{g}|$, $\Omega=-(g-\tilde{g})/2$.

\section{Addendum: Isotropic Heisenberg exchange $g(XX+YY+ZZ)$}
\label{sec:propertiesISOTROPIC}

Maximally entangling protocols introduced in previous sections 
are singular in the limit $\tilde{g}\rightarrow g$, which corresponds 
to the isotropic Heisenberg exchange interaction. Even though this 
limit is not met in superconducting qubits, for completeness, 
we briefly discuss it here.

It is obvious that when $g=\tilde{g}$,
the symmetric state $Y_{\pi/2}|000\ket$ is an eigenstate of 
the interaction Hamiltonian. Consequently, the Heisenberg exchange 
does not cause transitions out of it, making the gate time divergent. 
To perform single-step entanglement we break the symmetry 
of local rotations. For example, the GHZ state 
can be generated by
$
e^{-i\alpha} |{\rm GHZ}\ket =
e^{- i(\pi/2)\sigma_z^2}
e^{- i(\pi/3)\left(\sigma_y^1-\sigma_y^2\right)}
U_{\rm int}
e^{- i(\pi/12)\left(5\sigma_y^1+\sigma_y^2-3\sigma_y^3\right)}
e^{- i(\pi/2)\sigma_z^2}
|000\ket$,
with $\alpha = -\pi/2$, $t_{\rm GHZ} = (2/3)\times (\pi/2g)$.
To generate the W state, we generalize
Neeley's fast implementation for triangular $g(XX+YY)$ coupling 
\cite{NEELEY}
({\it cf.} \cite{MIGLIORE_Wstate}) 
to arbitrary coupling $g(XX+YY)+\tilde{g}ZZ$, including the 
Heisenberg exchange $g=\tilde{g}$:
$
e^{-i\alpha} |{\rm W}\ket = e^{+i(\pi/3)\sigma_z^2}
U_{\rm int}e^{- i(\pi/2)\sigma_y^2}|000\ket$,
with $\alpha = \left(5g-2\tilde{g}\right)\pi/18g$,
$t_{\rm W} = (4/9)\times (\pi/2g)$.

\section{Conclusion}

In summary, we have developed several single-step 
{\it symmetric} implementations for generating
maximally entangled tripartite quantum states in systems with 
anisotropic exchange interaction, which are directly applicable 
to superconducting qubit architectures.
In the GHZ case, both triangular and linear coupling schemes have 
been analyzed. 
In the isotropic limit, our implementations exhibit singularities
that can be removed by breaking the symmetry of 
the local pulses.


\begin{acknowledgments}

This work was supported by the DTO under
Grant No. W911NF-04-1-0204 and by the NSF under
Grants No. CMS-0404031 and No. CCF-0507227. The authors thank 
Michael Geller 
and
Matthew Neeley for helpful discussions.

\end{acknowledgments}

\end{document}